\documentclass[conference]{IEEEtran}
\IEEEoverridecommandlockouts
\usepackage{amsmath}
\usepackage{amsfonts}
\usepackage{bbding}
\usepackage{amssymb}
\usepackage{array}
\usepackage{subfigure}

\usepackage{graphicx}
\usepackage{subfigure}
\usepackage[named]{algo}
\usepackage{algorithmic}
\usepackage{psfrag}
\usepackage{stfloats}
\usepackage[compress]{cite}
\makeatletter
\renewcommand{\citepunct}{,\penalty\@m\hskip.13emplus.1emminus.1em}
\renewcommand{\citedash}{\hbox{--}\penalty\@m}
\makeatother
\usepackage{setspace}
\usepackage{color}
\allowdisplaybreaks

\usepackage{amsthm}
\usepackage{stfloats}

\newtheorem{pro}{Property}

\usepackage{hyperref}
\usepackage{bm}

\begin{document}
\title{Uplink Transmission Design with Massive Machine Type Devices in Tactile Internet}

\author{
\IEEEauthorblockN{{Changyang She and Chenyang Yang}} \vspace{0.0cm}
\IEEEauthorblockA{School of Electronics and Information
Engineering,\\ Beihang University, Beijing, China\\
Email:  \{cyshe,cyyang\}@buaa.edu.cn}\thanks{This work is supported partially by the National High Technology Research and Development Program of China under grant 2014AA01A703, National Basic Research Program of China, 973 Program under grant 2012CB316003 and National Natural Science Foundation of China (NSFC) under Grant 61120106002.}
\and \IEEEauthorblockN{{Tony Q. S. Quek}} \vspace{0.0cm}
\IEEEauthorblockA{$\quad\quad\quad\quad$Information Systems Technology and Design Pillar,\\ $\quad\quad\quad\quad\quad$Singapore University of Technology and Design, Singapore \\
Email: tonyquek@sutd.edu.sg} }

\maketitle
\begin{abstract} In this work, we study how to design uplink transmission with massive machine type devices in tactile internet, where ultra-short delay and ultra-high reliability are required. To characterize the transmission
reliability constraint, we employ a two-state transmission model based on the achievable rate with finite blocklength channel codes. If the channel gain exceeds a threshold, a short packet can be transmitted with a small error probability; otherwise there is a packet loss. To exploit frequency diversity, we assign multiple subchannels to each active device, from which
the device selects a subchannel with channel gain exceeding
the threshold for transmission. To show the total bandwidth required to ensure the reliability, we optimize the number of subchannels and bandwidth of each subchannel and the threshold for each device to minimize the total bandwidth of the system with a given number of antennas at the base station. Numerical results show that with $\bf{1000}$ devices in one cell, the required bandwidth of the optimized policy is acceptable even for prevalent cellular systems. Furthermore, we show that by increasing antennas at the BS, frequency diversity becomes unnecessary, and the required bandwidth is reduced.
\end{abstract}


\section{Introduction}
Achieving ultra-short end-to-end (E2E) delay and ultra-high reliability has become one of the major goals for the fifth generation (5G) cellular systems \cite{A2014Scenarios}. By ensuring ultra-low E2E delay with extremely small packet loss probability, {tactile internet} enables unprecedented mobile applications such as autonomous vehicles, mobile robots, augmented reality, and factory automation\cite{Gerhard2014The}.

Human-to-machine and machine-to-machine (M2M) communications are two major categories of application in tactile internet \cite{Meryem2016Tactile}, and the number of machine type devices is predicted to reach 26 billion by 2020 \cite{Erfan2016IoT}. Therefore, ensuring the stringent quality-of-service (QoS) for massive machine type devices becomes an urgent task in future wireless systems \cite{Popovski2014METIS}.
For human type communications, e.g., video conference, a certain amount of bandwidth can be reserved to each user. For machine type communications, however, a device may stay dumb for a long period between the transmissions of two subsequent short packets\cite{3GPP2012MTC}. Thus, it is not necessary to reserve bandwidth for each device. Moreover, in the scenarios that the number of devices is large, reserving bandwidth for each device leads to unaffordable total bandwidth requirement. To save bandwidth, the base station (BS) can only assign bandwidth to the active devices that require to transmit data.

Recently, the problem on how to ensure ultra-low E2E delay and ultra-high reliability has drawn increasing attention from academic and industrial communities \cite{Adnan2015Towards,David2014Achieving,Martin2015Channel,Beatriz2015Reliable}. Resource allocation for tactile internet applications has been studied in \cite{Adnan2015Towards}, where the queueing delay and queueing delay violation probability are taken into account. Studies in \cite{David2014Achieving,Martin2015Channel,Beatriz2015Reliable} show how to exploit diversity to increase reliability. In these works, the instantaneous data rate is characterized by the Shannon capacity, which is applicable when the blocklength of channel codes goes to infinite. Under ultra-short delay constraint, the blocklength is finite, and the Shannon capacity is not achievable. As a consequence, the ultra-low packet loss probability cannot be ensured.

The achievable rate with finite blocklength channel codes obtained in \cite{Yury2010Channel} has been applied to analyze queueing delay for real-time services and tactile internet in downlink transmission \cite{Gross2015Delay,She2016GC}. For real-time service with queueing delay around $5 \sim 10$~ms, if Shannon capacity is used to design transmission policy, then the queueing delay bound and the delay bound requirement cannot be satisfied \cite{Gross2015Delay}. Based on this observation, the achievable rate in finite blocklength regimes is applied in downlink transmission design for tactile internet \cite{She2016GC}, where a short time (i.e., half of a short frame) is reserved for uplink (UL) transmission. However, the question on how to design UL transmission policy under ultra-short delay and ultra-high reliability remains open.

In this work, we focus on UL transmission design for tactile
internet. We investigate the impact of spatial diversity and frequency diversity on ensuring the transmission reliability, and the total bandwidth required for a wireless system to support the QoS requirement of massive machine type devices. To this end, we employ a two-state transmission model to
characterize the transmission reliability constraint based on the achievable rate with finite blocklength channel codes. We assign multiple subchannels to each active device, from which the device simply selects one subchannel with channel power exceeding a threshold  for transmission after channel probing \cite{Johnston2015The}. We optimize the number of subchannels, the bandwidth of each subchannel, and the threshold for each device to minimize the total bandwidth required by the system to ensure the reliability. Numerical results are provided to show the required total bandwidth and the impact of diversity.

\section{System Model}
Consider a cellular system, where a BS with $N_\mathrm{t}$ antennas is accessed by $M$ single-antenna devices. The packets generated at the devices are transmitted to the BS. After receiving the packets successfully, the BS sends the packets to related destinations. Before sending a transmission request, a device should first access to the BS \cite{Yuan2016IoT}. Assume that the devices stay in one cell, and all the devices have accessed to the BS.

In this work, we focus on the UL transmission design for massive machine
type devices, where the bandwidth will be assigned to a device only when it has a packet to transmit. To exploit frequency diversity, the BS assigns $N_m$ subchannels with independent channel gains to the $m$th device, as detailed later. Since the interference among devices causes severe deterioration in QoS, we assume that different subchannels are used for the devices requesting for concurrent transmissions.

The QoS provision is characterized by an E2E delay $D_{\max}$ imposed on each packet of a device and the overall reliability $\varepsilon_{\max}$ required by the device. In LTE systems, the E2E delay may include UL and downlink transmission delay, coding and processing delay,
queueing delay, and routing delay in backhaul and core networks, while the overall reliability may include the packet loss due to transmission error and queueing delay violation. In this work, we restrict to the UL  transmission delay and reliability. Specifically, the UL transmission procedure for each packet should be completed within a short time  $D^{\rm U} < D_{\max}$ with a transmission error induced packet loss probability $\varepsilon ^{\rm U} < \varepsilon_{\max}$ for each device.

To reduce transmission delay, we consider the short frame structure proposed in \cite{Petteri2015A}, where time is discretized into frames with duration $T_\mathrm{f}$, which equals to the  transmission time interval (TTI) of the system.

The UL transmission procedure includes the following steps: (i)  generation of a packet by a device; (ii) UL transmission request from the device; (iii) bandwidth assignment at the BS; (iv) transmission grant and channel probing; (v) subchannel selection and packet transmission by the device.
By assuming negligible processing delay, the delay caused by the control signaling and data transmission is $D^{\rm U} = 3T_{\rm f}$ \cite{Shehzad2015Control}.

Then, the transmission time of each packet should be less than the duration of one frame, $T_{\rm f}$. This implies that retransmission in subsequent frames is not allowed  in the case of a packet with error. In fact, retransmitting a packet can hardly improve the successful transmission probability when the
channels in multiple frames stay in deep fading.

\subsection{Traffic Model}
As shown in \cite{3GPP2012MTC,Mehdi2013Performance}, the packet arrival process in vehicle networks as well as in some other M2M communications can be modeled as Poisson processes, which is an aggregation of packets generated by multiple devices, where the packets generated by each device follow a Bernoulli process. Specifically, in each frame a device either transmits one packet or stays dumb randomly. Each packet includes $u$~bits information, which is small in machine type communications.


\subsection{Channel Probing for Subchannel Selection}
For the device with low maximal transmit power, equally allocating the power over multiple subchannels leads to low signal-to-noise ratio (SNR). If a device knows the channel gains of the assigned subchannels, it can select several subchannels with high gains to transmit. Nevertheless, for the machine type devices with small packets, a single subchannel in good condition is sufficient to transmit a packet with small error probability. This suggests that a device only needs to find the  subchannels whose channel gains exceed a certain threshold, from which the device can randomly select one to transmit the packet.

To help each active device, say the $m$th device, determine whether the instantaneous channel gain of a subchannel is above the threshold $g_m^{\rm{th}}$ or not, the BS only needs to broadcast downlink pilots that are orthogonal among subchannels. This procedure is referred to as \emph{channel probing} as in \cite{Johnston2015The}.

\subsection{Channel Model}
Without retransmission, the UL data transmission for each packet should be completed in
one frame. For the devices with low and medium velocity, the channel coherence time $T_{\rm c}$ is much longer than the frame duration. To ensure low packet loss probability with the stringent delay requirement, channel coding should be performed within each frame, during which the channel is static. This kind of channel is referred to as \emph{quasi-static fading channel} in \cite{Yury2014Quasi}.

\begin{figure}[htbp]
        \vspace{-0.1cm}
        \centering
        \begin{minipage}[t]{0.38\textwidth}
        \includegraphics[width=1\textwidth]{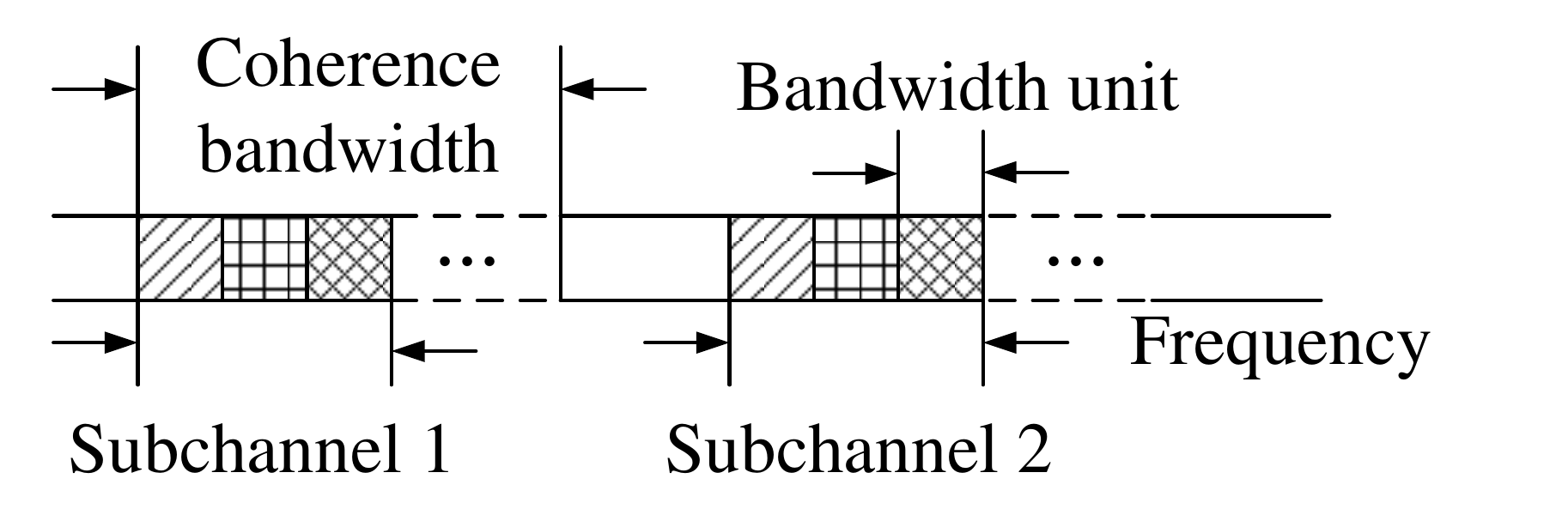}
        \end{minipage}
        \vspace{-0.2cm}
        \caption{Illustration of bandwidth allocation.}
        \label{fig:frequency}
        \vspace{-0.2cm}
\end{figure}

Consider a frequency-selective channel.
To maximize frequency diversity gain, the instantaneous channel gains on the $N_m$ subchannels assigned to the $m$th device should be
independent, and hence the frequency separation of adjacent subchannels should be larger than the channel coherence bandwidth  $W_{\rm c}$, as  shown in Fig. \ref{fig:frequency}. In real-world systems, frequency is discretized into basic bandwidth units, e.g., subcarriers in orthogonal frequency division multiple access systems. Each subchannel consists of multiple bandwidth units. By adjusting the number of bandwidth units in one subchannel, the bandwidth of each subchannel can be changed. Denote $B_m$ as the bandwidth of each subchannel assigned to the $m$th device. We assume that $B_m<W_{\rm c}$, such that each subchannel is subject to frequency-flat fading.

Denote the average channel gain of the $m$th device as $\alpha_m$, and the instantaneous channel gain on the $i$th subchannel assigned to the $m$th device as $g_{m,i}$. After channel probing, the device selects a subchannel with estimated instantaneous channel gain $g^{+}_{m} = ({{\bf{h}}_m})^H{\bf{h}}_m$, where  $[ \cdot ]^H$ denotes the conjugate transpose and ${\bf{h}}_{m}$ is the channel vector whose elements are independent and identically complex Gaussian distributed with zero mean and unit variance. Assume that both $\alpha_m$ and ${\bf{h}}_{m}$ are perfectly known at the BS.
Since the machine type devices are usually of low cost, it is reasonable to assume that each active device simply employs the maximal transmit power to transmit the packet. For a single-input-multiple-output system, the maximal number of bits that can be transmitted from the $m$th device to the BS in one frame can be accurately approximated as \cite{Yury2014Quasi}
\begin{align}
s_m \approx \frac{ T_{\rm f} B_m}{\ln{2}} \left\{\ln\left(1+\frac{\alpha_m P^{\max} g^{+}_m}{ N_0 B_m}\right) - \sqrt{\frac{V_m}{T_{\rm f} B_m}}f_{\rm Q}^{-1}({\varepsilon_m })\right\}, \label{eq:sn}
\end{align}
where $P^{\max}$ is the maximal transmit power of each device, $N_0$ is the single-sided noise spectral density, $\varepsilon_m$ is the transmission error probability (i.e.,
the block error probability) of the $m$th device, $f_{\mathrm Q}^{-1}(x)$ is the inverse of the Q-function,  and $V_m = 1-{\left[1+\frac{\alpha_m P^{\max} g^+_m}{ N_0 B_m}\right]^{-2}}$ \cite{Yury2014Quasi}.

The number of symbols transmitted in one frame, also referred to as blocklength, is determined by the bandwidth and transmission duration according to $l_m = T_{\rm f} B_m$. For large $l_m$, \eqref{eq:sn} can be approximated by the Shannon capacity, i.e.
\begin{align}
s^{\infty}_m =  \frac{ T_{\rm f} B_m}{\ln{2}}\ln\left(1+\frac{\alpha_m P^{\max} g^{+}_m}{ N_0 B_m}\right).\label{eq:Shannon}
\end{align}

\section{Transmission Constraint on Reliability }

\subsection{Two-state Transmission Model}
To analyze the availability or reliability of tactile internet, a transmission model based on the Shannon capacity was applied in existing studies \cite{David2014Achieving,Martin2015Channel}. Specifically, if the received SNR is higher than a threshold, a packet with size $u$ can  be transmitted successfully, i.e., $s_m^{\infty} \geq u$. Otherwise, an outage occurs. However, with finite blocklength channel codes, the transmission error probability $\varepsilon_m$ in \eqref{eq:sn} is always positive such that $s_m< s_m^{\infty} $, and cannot be ignored under ultra-high reliability requirement. Therefore, the existing transmission model underestimates the transmission error probability in high SNR regimes, and hence cannot ensure ultra-high reliability.

Further recalling that retransmission is not allowed, we consider a transmission model based on the achievable rate with finite blocklength channel codes in \eqref{eq:sn}. If the instantaneous channel gain $g_m$ is higher than a threshold $g_m^{\rm{th}}$ such that $s_m\geq u$, then the packet can be transmitted successfully with probability $1 - \varepsilon_m$. Otherwise, there is a packet loss. Substituting \eqref{eq:sn} into $s_m = u$, the threshold can be accurately approximated as
\begin{align}
g_m^{\rm{th}} \approx \frac{{{N_0}{B_m}}}{{{\alpha _m}{P^{\max}}}}\left\{ {\exp \left[ {\frac{{u\ln 2}}{{T_{\rm f} {B_m}}} + \sqrt {\frac{1}{{T_{\rm f} {B_m}}}} f_{\rm Q}^{ - 1}\left( {\varepsilon _{m} } \right)} \right] - 1} \right\},\label{eq:thresh}
\end{align}
since $V_m \approx 1$ is accurate in high SNR regime.

Such a transmission model considers the transmission error probability when $g^+_m \geq g_m^{\rm{th}}$, and hence is applicable for designing transmit policy under the ultra-high reliability constraint. Since the model depends on two states of the channel, we refer to it as a \emph{two-state transmission model}.

\subsection{Constraint on UL Transmission Reliability}

For the UL transmission of the $m$th device, the probability that there is at least one subchannel with instantaneous channel gain higher than $g_m^{\rm {th}}$ can be expressed as follows:
\begin{align}
\Pr \left\{ \mathop  \cup \limits_{i = 1}^{N_m} \left\{ {{g_{m,i}} \ge g_m^{\rm{th}}} \right\} \right\} &= 1 - \prod\limits_{i = 1}^{N_m} {\Pr \left\{ {{g_{m,i}} < g_m^{{\rm{th}}}} \right\}} \nonumber\\
&= 1 - {\left[ {\int_0^{g_m^{{\rm{th}}}} {{f_{\rm{g}}}\left( x \right)dx} } \right]^{N_m}}\label{eq:Prg},
\end{align}
where ${f_{\rm{g}}}\left( x \right)$ is the distribution of instantaneous channel gain. Since the elements of ${\bf{h}}_m$ are complex Gaussian distributed, we have $f_{\rm g}\left( x \right) = \frac{1}{{\left( {{N_\mathrm{t}} - 1} \right)!}}{x^{{N_\mathrm{t}} - 1}}{e^{ - x}}$.

We use an indicator function to represent whether the packet is successfully transmitted or not. If the packet is successfully transmitted to the BS from the $m$th device, then ${\bf{1}}_m = 1$. Otherwise, ${\bf{1}}_m = 0$. From \eqref{eq:Prg}, we have
\begin{align}
\Pr\{{\bf{1}}_m = 1\} &= \Pr \left\{ \mathop  \cup \limits_{i = 1}^{N_m} \left\{ {{g_{m,i}} \ge g_m^{\rm{th}}} \right\} \right\}(1-\varepsilon _m ) \nonumber\\
&\approx 1 - {\left[ {\int_0^{g^{\rm{th}}_m} {{f_{\rm{g}}}\left( x \right)dx} } \right]^{N_m}} - \varepsilon_m .\label{eq:PrA}
\end{align}
The above approximation is accurate since ${\left[ {\int_0^{g_m^{{\rm{th}}}} {{f_{\rm{g}}}\left( x \right)dx} } \right]^{N_m}}$ and $\varepsilon _m $ are extremely small. To ensure the transmission reliability in UL, the following constraint should be satisfied,
\begin{align}
f_{\rm u}(N_m,B_m, \varepsilon_m) \triangleq {\left[ {\int_0^{g^{\rm{th}}_m} {{f_{\rm{g}}}\left( x \right)dx} } \right]^{N_m}} + \varepsilon_m  \leq \varepsilon^{\mathrm U}. \label{eq:relia}
\end{align}

\section{Transmit Policy Optimization}
In this section, we optimize the transmit policy that minimizes the required bandwidth to satisfy the QoS requirement of massive machine type devices.

To facilitate channel probing in selecting one subchannel, the DL pilot overhead linearly increases with $N_m$. To make the overhead  acceptable, $N_m$ cannot be too large. Denote the maximal number of subchannels that can be assigned to each device as $N_{\max}$. Then, we have $N_m \leq N_{\max}$.

The threshold $g_m^{\rm th}$ in \eqref{eq:thresh} depends on $B_m$ and $\varepsilon _m$. Therefore, with given $B_m$, the threshold can be adjusted by controlling the value of $\varepsilon _m$. To determine how much bandwidth is required to ensure the reliability with a large number of devices, we optimize the values of $N_m$, ${B_m}$, and $\varepsilon _m $ that minimizes the overall bandwidth from the following problem:
\begin{align}
\mathop {\mathop {\min}\limits_{N_m,{B_m},\varepsilon _m } }\limits_{m = 1,...,{M_{\rm a}}} & \sum\limits_{m = 1}^{{M_{\rm a}}} {N_m}{B_m} \label{eq:Wtot}\\
\text{s.t.}\;& 0 < {B_m} \le {W_{\rm c}},\label{eq:Wc}\tag{\theequation a} \\
& 0 < N_m \le N_{\max}, N_m \in {\mathbb{Z}}, \label{eq:Nh}\tag{\theequation b}\\
& 0 < \varepsilon_m  < \varepsilon^{\rm U}, \label{eq:epson}\tag{\theequation c}\\
& f_{\rm u}(N_m,B_m, \varepsilon_m) \leq \varepsilon^{\mathrm U}, \label{eq:fe}\tag{\theequation d}
\end{align}
where $M_{\rm a}$ is the number of active devices that need to transmit packets in a frame.

In the sequel, we propose an algorithm to find the global optimal solution of problem \eqref{eq:Wtot}. Since the constraints for each device does not depend on those of the other devices, problem \eqref{eq:Wtot} can be equivalently decomposed into $M_{\rm a}$ single-device problems as follows:
\begin{align}
\mathop {\mathop {\min}\limits_{N_m,{B_m},\varepsilon _m } } & {N_m{B_m}} \label{eq:su}\\
\text{s.t.} & \;\eqref{eq:Wc}, \eqref{eq:Nh}, \eqref{eq:epson}\;\text{and}\;\eqref{eq:fe}.\nonumber
\end{align}

To solve problem \eqref{eq:su}, we need some properties of $f_{\rm u}(N_m,B_m, \varepsilon_m)$.
\begin{pro}\label{P:fw}
\emph{Given the values of $N_m$ and $\varepsilon_m$, $f_{\rm u}(N_m,B_m, \varepsilon_m)$ strictly decreases with $B_m$.}
\begin{proof}
See Appendix \ref{App:Appendix_P1}.
\end{proof}
\end{pro}

Based on Property \ref{P:fw}, we have the following property,
\begin{pro}\label{P:w}
\emph{Given the value of $N_m$, $f_{\rm u}(N_m,B_m, \varepsilon^*_m(B_m))$ strictly decreases with $B_m$, where $\varepsilon^*_m(B_m)$ is the optimal value of $\varepsilon_m$ that minimizes $f_{\rm u}(N_m,B_m, \varepsilon_m)$ with given $B_m$.}
\begin{proof}
See Appendix \ref{App:Appendix_P2}.
\end{proof}
\end{pro}

Since the value of $f_{\rm u}(N_m,B_m, \varepsilon^*_m(B_m))$ with given $N_m$ strictly decreases with $B_m$, the minimal required $B_m$ that satisfies \eqref{eq:relia} with given $N_m$ can be obtained when $f_{\rm u}(N_m,B_m, \varepsilon^*_m(B_m)) = \varepsilon _{\mathrm U}$, and hence the minimal $B_m$ can be obtained via the binary search method \cite{boyd}. The searching algorithm needs to compute the value of $f_{\rm u}(N_m,B_m, \varepsilon^*_m(B_m))$, and hence needs to find $\varepsilon^*_m(B_m)$ with given $B_m$. To show when $\varepsilon^*_m(B_m)$ can be obtained with a low complexity method, we need the following property.

\begin{pro}\label{P:fe}
\emph{Given the values of $N_m$ and $B_m$, $f_{\rm u}(N_m,B_m, \varepsilon_m)$ is convex in $\varepsilon_m$ when $g^{\rm{th}}_m < N_{\rm t}-1$.}
\begin{proof}
See proof in Appendix \ref{App:Appendix_P3}.
\end{proof}
\end{pro}

If $f_{\rm u}(N_m,B_m, \varepsilon_m)$ is convex in $\varepsilon_m$, e.g., $g^{\rm th}_m < N_{\rm t}-1$, then the global optimal solution $\varepsilon^*_m(B_m)$ can be obtained by the exact linear search method \cite{boyd}. Otherwise, to obtain the global optimal $\varepsilon_m(B_m)$, the exhaustive search method should be used. Note that to ensure ultra-high reliability in \eqref{eq:relia}, $g^{\rm th}_m$ cannot be too large. For example, when $N_{\rm t} \geq 2$ and $\varepsilon_{\max} \leq 10^{-5}$, which is true for most tactile internet applications, we have $g^{\rm{th}}_m < N_{\rm t}-1$ under constraint \eqref{eq:relia} in the cases $N_{\max} \leq 10$.

Based on Property \ref{P:w} and Property \ref{P:fe}, we propose a searching algorithm to find the optimal solution of problem \eqref{eq:Wtot}. Given the value of $N_m$, the optimal values of $B_m$ and $\varepsilon_m$ that minimizes \eqref{eq:su} can be found via the binary search method and the exact linear search method. By searching $B_m$ and $\varepsilon_m$ with different values of $N_m \in \{ 1,...,N_{\max}\}$, the optimal solution of problem \eqref{eq:su} can be obtained, and is denoted as $\{N^{*}_m, B^*_m, \varepsilon^{*}_m \}$. To find the optimal solution of the original problem \eqref{eq:Wtot}, the system needs to solve problem \eqref{eq:su} $M_{\rm a}$ times. Hence, the complexity of problem \eqref{eq:Wtot} is $O(M_{\rm a} N_{\max})$, which linearly increases with the number of active devices. The details of the algorithm are provided in Table I.

\vspace{-0.2cm}
\renewcommand{\algorithmicrequire}{\textbf{Input:}}
\renewcommand{\algorithmicensure}{\textbf{Output:}}
\begin{table}[htb]\small
\caption{Algorithm to Find the Global Optimal Solution of Problem \eqref{eq:Wtot}}
\vspace{-0.6cm}
\begin{tabular}{p{8.5cm}}
\\\hline
\end{tabular}
\vspace{-0.2cm}
\begin{algorithmic}[1]
\REQUIRE $M_{\rm a}$, $N_{\max}$, $T_{\rm f}$, $u$, $N_0$, $N_\mathrm{t}$, $\alpha_m$, $P^{\max}$, and accuracy requirement of the binary search method $\delta_b$.
\ENSURE $N^{*}_m$, $B^*_m$, and $\varepsilon^{*}_m$, $m=1,...,M_{\rm a}$.
\STATE Set $m := 1$.
\WHILE{$m \leq M_{\rm a}$}
\STATE $N_m := 1$
\WHILE{$N_m \leq N_{\max}$}
\STATE Set $B_{\rm lb} := 0$, $B_{\rm ub} := W_{\rm c}$, $B_{\rm bs} := 0.5(B_{\rm lb}+B_{\rm ub})$.
\WHILE{$B_{\rm ub} - B_{\rm lb} > \delta_b$}
\STATE Apply the exact linear search method to find $\varepsilon_{\rm bs}$ that minimizes $f_{\rm u}(N_m,B_{\rm bs}, \varepsilon_{\rm bs})$.
\IF{$f_{\rm u}(N_m,B_{\rm bs}, \varepsilon_{\rm bs}) > \varepsilon^{\rm U}$}
\STATE $B_{\rm lb} := B_{\rm bs}$, $B_{\rm bs} := 0.5(B_{\rm lb}+B_{\rm ub})$.
\ELSE
\STATE $B_{\rm ub} := B_{\rm bs}$, $B_{\rm bs} := 0.5(B_{\rm lb}+B_{\rm ub})$.
\ENDIF
\ENDWHILE
\IF{$f_{\rm u}(N_m,B_{\rm bs}, \varepsilon_{\rm bs}) \leq \varepsilon^{\rm U}$}
\STATE $B_m(N_m) := B_{\rm bs}$ and $\varepsilon_m(N_m):=\varepsilon_{\rm bs}$.
\ELSE
\STATE $B_m(N_m) := \text{NaN}$ and $\varepsilon_m(N_m):= \text{NaN}$.
\ENDIF
\ENDWHILE
\STATE $N_m^{*} := \mathop {\arg }\limits_{N_m} \min N_m{B_m}\left( {N_m} \right)$.
\STATE $B^*_m := B_m(N_m^{*})$, $\varepsilon_m^* := \varepsilon_m(N_m^{*})$.
\ENDWHILE
\RETURN $N_m^{*}, B^*_m, \varepsilon_m^*$, $m = 1,...,M_{\rm a}$.
\end{algorithmic}
\vspace{-0.2cm}
\begin{tabular}{p{8.5cm}}
\\
\hline
\end{tabular}
\vspace{-0.5cm}
\end{table}

\section{Numerical Results}

In this section, we demonstrate the required radio resources to support the stringent uplink transmission reliability. To observe the impact of different factors, we first consider the channel only with path loss, and then extend to more realistic channel with log normal  shadowing.

The number of devices in one cell is $M=1000$. The distances between devices and the BS are uniformly distributed in $[50,250]$~m. The path loss model is $-10\lg(\alpha_m)=35.3+37.6 \lg(d_m)$, where $d_m$ is the distance between the $m$th device and the BS. The maximal transmit power of each device is set to be $23$~dBm ($P^{\max} = 0.2$~W). Each packet contains $20$~bytes data. The reliability requirement is $\varepsilon _{\max} = 1-99.99999\%$ \cite{Gerhard2014The}, and $\varepsilon^{\rm U} = 0.5 \varepsilon _{\max}$. The single-sided noise spectral density and coherence bandwidth is set to be $N_0 = -174$~dBm/Hz and $W_{\rm c} = 0.5$~MHz, respectively. The frame duration is $T_f = 0.1$~ms. The maximal number of subchannels assigned to each active device is set to be $N_{\max} = 10$. This setup is used for all following results unless otherwise specified.

\begin{figure}[htbp]
        \vspace{-0.4cm}
        \centering
        \begin{minipage}[t]{0.45\textwidth}
        \includegraphics[width=1\textwidth]{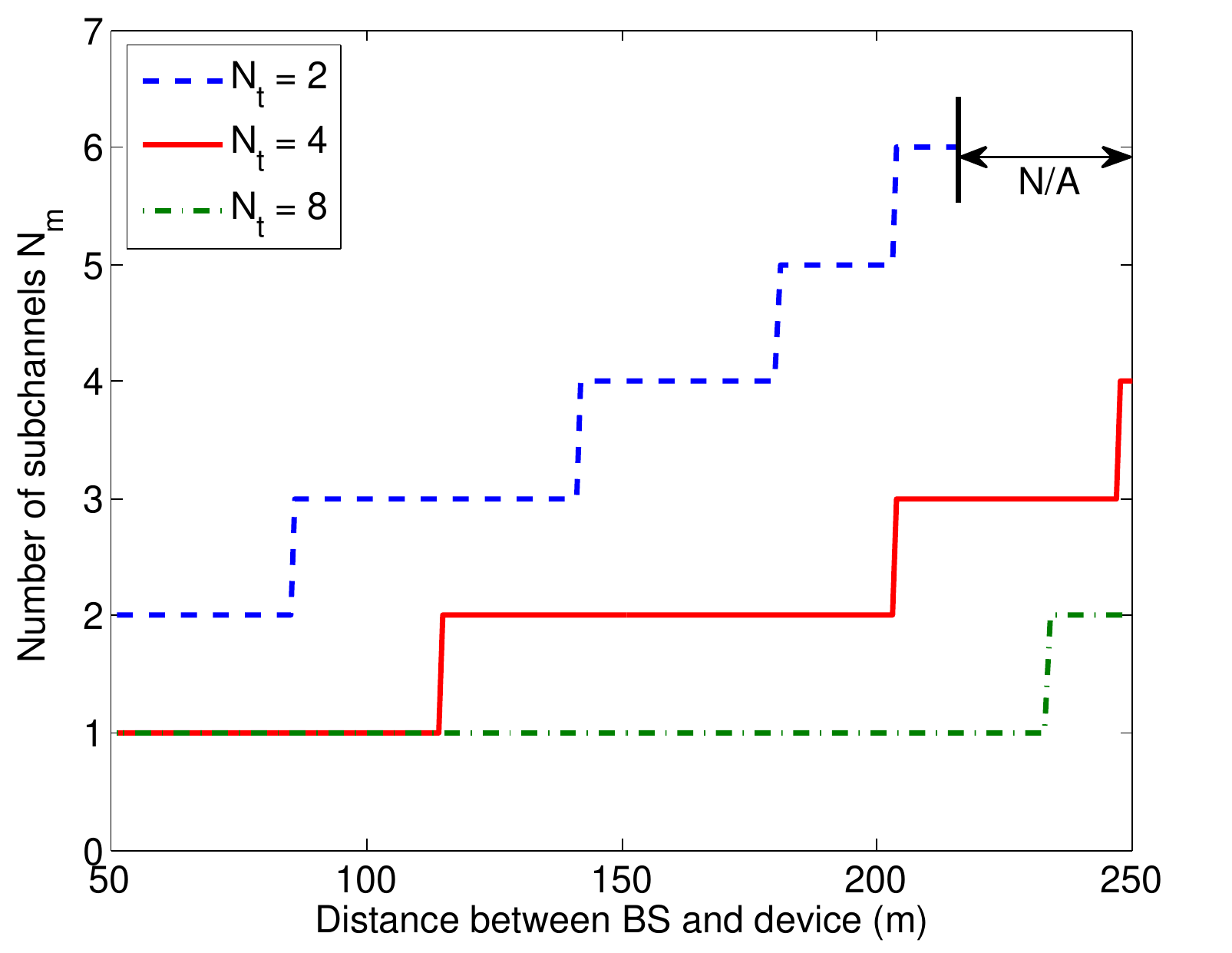}
        \end{minipage}
        \vspace{-0.4cm}
        \caption{Optimal number of links for frequency diversity.}
        \label{fig:Ns}
        \vspace{-0.3cm}
\end{figure}

The optimal number of subchannels assigned to each active device with different device-BS distances is shown in Fig. \ref{fig:Ns}. If the BS is only equipped with two antennas ($N_{\rm t} = 2$), the QoS requirement of some devices at the cell edge cannot be satisfied even when $B_m = W_{\rm c}$ and $N_m = N_{\max}$, i.e. problem \eqref{eq:Wtot} is infeasible. When $N_{\rm t} \geq 16$, $N^*_m = 1$ for all the devices (which is not shown in the figure), i.e., only one subchannel is assigned to each device. In other words, with large spatial diversity, frequency diversity is unnecessary.

\begin{figure}[htbp]
        \vspace{-0.4cm}
        \centering
        \begin{minipage}[t]{0.45\textwidth}
        \includegraphics[width=1\textwidth]{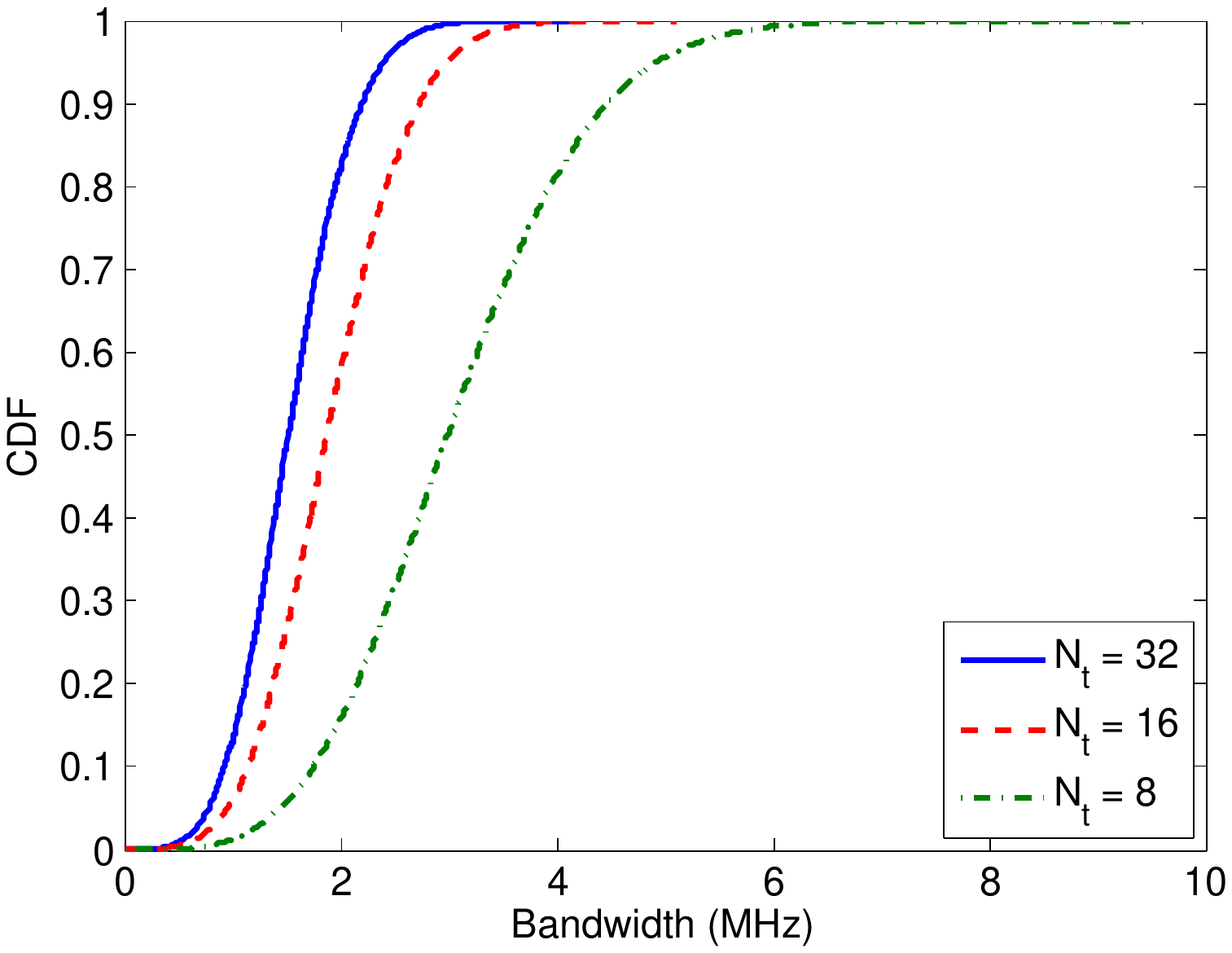}
        \end{minipage}
        \vspace{-0.4cm}
        \caption{CDF of total bandwidth required to support the QoS.}
        \label{fig:Wreq}
        \vspace{-0.3cm}
\end{figure}
The cumulative distribution function (CDF) of the total bandwidth of the system required to support the QoS with the optimized transmit policy is shown in Fig. \ref{fig:Wreq}. The CDF is obtained by calculating the total required bandwidth in $10^6$~frames. The average packet rate is set to be $100$~packets/s, which is relatively high in vehicle communications and M2M communications \cite{Mehdi2013Performance,3GPP2012MTC}. With frame duration $T_f = 0.1$~ms, the average packet rate is one packet per $100$~frames. In other words, during $99$\% of time, a device stays dumb. If the system reserves bandwidth for each device like human type communications, the required total bandwidth with $32$ receive antennas is $155$~MHz, which is obtained from solving problem \eqref{eq:Wtot} by setting $N_t=32$ and $M_a=1000$. With the optimized policy, the results show that the required total bandwidth for $N_{\rm t} \geq 8$ is less than $10$~MHz, because the BS only assigns bandwidth to the active devices that require to transmit packets in each frame.

In what follows, we take into account shadowing, which follows a log normal distribution with zero mean and $8$~dB standard deviation \cite{3GPP2010Shadowing}, i.e., $-10\lg(\alpha_m)=35.3+37.6 \lg(d_m) + \delta^s_m$, where $\delta^s_m \sim {\mathcal{N}}(0,8)$. With shadowing, the percentage of devices that the QoS in \eqref{eq:relia} can not be satisfied is listed in Table II, which is obtained by solving problem \eqref{eq:Wtot}. The results show that the percentage of devices without QoS guarantee decreases slowly by increasing receive antennas at the BS, owing to the array gain. However, the QoS requirement of some devices still cannot be satisfied even with large $N_{\rm t}$. While macro diversity can deal with such a problem to some extent \cite{Beatriz2015Reliable}, other more effective solutions are in urgent need.

\begin{table}[btp]
\vspace{-0.2cm}
\small
\renewcommand{\arraystretch}{1.3}
\caption{Percentage of devices without QoS guarantee}
\begin{center}\vspace{-0.2cm}
\begin{tabular}{|p{2cm}|p{1cm}|p{1cm}|p{1cm}|p{1cm}|}
\hline
  $N_{\rm t}$ & $16$ & $32$ & $64$ & 128 \\\hline
  Percentage & $6.2$\% & $2.8$\%& $1.5$\% & $1.0$\%  \\\hline
\end{tabular}
\end{center}
\vspace{-0.6cm}
\end{table}

\section{Conclusion}
In this paper, we studied uplink transmission optimization for massive machine type devices in tactile internet. We employed a two-state transmission model using the achievable rate with finite blocklength channel codes to reflect the reliability constraint. Then, we proposed an algorithm to find the optimal number of links for frequency diversity, optimal bandwidth and transmission threshold required to ensure the uplink transmission reliability for each active device that minimizes the total bandwidth of the system. Numerical results validated that the total bandwidth required by the optimized policy is much smaller than reserving bandwidth for each device like human type
communications, which is acceptable even for the prevalent cellular systems. The results showed that by increasing the number of antennas at the BS, the required total bandwidth can be reduced, and the percentage of devices without QoS guarantee due to shadowing decreases.

\appendices
\section{Proof of Property \ref{P:fw}}
\label{App:Appendix_P1}
\renewcommand{\theequation}{A.\arabic{equation}}
\setcounter{equation}{0}
\begin{proof}
By substituting $f_{\rm g}\left( x \right) = \frac{1}{{\left( {{N_\mathrm{t}} - 1} \right)!}}{x^{{N_\mathrm{t}} - 1}}{e^{ - x}}$ into \eqref{eq:relia}, we have
\begin{align}
{f_{\rm{u}}}(N_m,{B_m},{\varepsilon _m}) = {\left[ {\int_0^{g_m^{{\rm{th}}}} {\frac{1}{{\left( {{N_{\rm{t}}} - 1} \right)!}}{x^{{N_{\rm{t}}} - 1}}{e^{ - x}}dx} } \right]^{N_m}} + {\varepsilon _m}.\label{eq:feA}
\end{align}
Denote ${f_{\rm{e}}} = {\int_0^{g_m^{{\rm{th}}}} {\frac{1}{{\left( {{N_{\rm{t}}} - 1} \right)!}}{x^{{N_{\rm{t}}} - 1}}{e^{ - x}}dx} }$. To prove ${f_{\rm{u}}}(N_m,{B_m},{\varepsilon _m})$ strictly decreases with ${B_m}$, we only need to prove that $\left({f_{\rm{e}}}\right)^{N_m}$ strictly decreases with ${B _m}$. To this end, we first prove that $\left(f_{\rm{e}}\right)^{{N_m}}$ strictly increases with $g_m^{\rm th}$ and then prove that $g_m^{\rm th}$ strictly decreases with $B_m$. From \eqref{eq:feA}, we can obtain that
\begin{align}
\frac{{\partial {\left(f_{{\rm{e}}}\right)^{N_m}}}}{{\partial g_m^{{\rm{th}}}}} = {\left(f_{{\rm{e}}} \right)^{N_m - 1}}\frac{{N_m{{\left( {g_m^{{\rm{th}}}} \right)}^{{N_{\rm{t}}} - 1}}{e^{ - g_m^{{\rm{th}}}}}}}{{\left( {{N_{\rm{t}}} - 1} \right)!}} > 0.\label{eq:dfdg}
\end{align}
As a result, $\left(f_{\rm{e}}\right)^{{N_m}}$ strictly increases with $g_m^{\rm th}$. Denote $x = \sqrt{\frac{1}{B_m}}$. Then, \eqref{eq:thresh} can be rewritten as follows,
\begin{align}
g_m^{\rm th} = \frac{C_1}{x^2}\left[\exp\left(C_2x^2+C_3x\right)-1\right],\label{eq:gxA}
\end{align}
where $C_1 = \frac{N_0}{\alpha_m P^{\max}} > 0$, $C_2 = \frac{u \ln 2}{T_{\rm f}}>0$ and $C_3 = \sqrt {\frac{1}{{T_{\rm f}}}} f_{\rm Q}^{ - 1}\left( {\varepsilon _m } \right) > 0$. From \eqref{eq:gxA}, we can derive that
\begin{align}
\frac{{dg_m^{{\rm{th}}}}}{{dx}} = \frac{{{C_1}}}{{{x^3}}}\left[ {{e^{{C_2}{x^2} + {C_3}x}}\left( {2{C_2}{x^2} + {C_3}x - 1} \right) + 1} \right].\label{eq:dgdxA}
\end{align}
To prove $\frac{{dg_m^{{\rm{th}}}}}{{dx}}>0$, we only need to show that $\tilde{f}(y) = e^y(y-1)+1 > 0$, where $y = {C_2}{x^2} + {C_3}x \in (0,\infty)$. Since $\mathop {\lim }\limits_{y \to 0} \tilde{f}(y) = 0$ and $\tilde{f}'(y) = ye^y>0$, $\tilde{f}(y) = e^y(y-1)+1 > 0, \forall y \in (0,\infty)$. Therefore, $\frac{{dg_m^{{\rm{th}}}}}{{dx}} > 0$. Moreover, $x = \sqrt{\frac{1}{B_m}}$, which strictly decreases with $B_m$. Hence, $g_m^{{\rm{th}}}$ strictly decreases with $B_m$. This completes the proof.
\end{proof}

\section{Proof of Property \ref{P:w}}
\label{App:Appendix_P2}
\renewcommand{\theequation}{B.\arabic{equation}}
\setcounter{equation}{0}
\begin{proof}
To prove that $f_{\rm u}(N_m,B_m, \varepsilon^*_m(B_m))$ decreases with $B_m$, we show that for any $W_m < B_m$, $f_{\rm u}(N_m,W_m, \varepsilon^*_m(W_m)) > f_{\rm u}(N_m,B_m, \varepsilon^*_m(B_m))$. According to Property \ref{P:fw}, given $N_m$ and $\varepsilon^*_m(W_m)$, we have
\begin{align}
f_{\rm u}(N_m,W_m, \varepsilon^*_m(W_m)) > f_{\rm u}(N_m,B_m, \varepsilon^*_m(W_m)).\label{eq:feB1}
\end{align}
Since $\varepsilon^*_m(B_m)$ is the optimal value of $\varepsilon_m$ that minimizes $f_{\rm u}(N_m,B_m, \varepsilon_m)$, we have
\begin{align}
f_{\rm u}(N_m,B_m, \varepsilon^*_m(W_m)) \geq f_{\rm u}(N_m,B_m, \varepsilon^*_m(B_m)).\label{eq:feB2}
\end{align}
From \eqref{eq:feB1} and \eqref{eq:feB2}, we can obtain that
\begin{align}
f_{\rm u}(N_m,W_m, \varepsilon^*_m(W_m)) > f_{\rm u}(N_m,B_m, \varepsilon^*_m(B_m))\nonumber.
\end{align}
The proof follows.
\end{proof}

\section{Proof of Property \ref{P:fe}}
\label{App:Appendix_P3}
\renewcommand{\theequation}{C.\arabic{equation}}
\setcounter{equation}{0}
\begin{proof}
According to \eqref{eq:feA}, to study the convexity of ${f_{\rm{u}}}(N_m,{B_m},{\varepsilon _m})$, we only need to study the convexity of $\left({f_{\rm{e}}}\right)^{N_m}$. To this end, we first prove that $g_m^{\rm th}$ in \eqref{eq:thresh} is convex in ${\varepsilon _m}$. Then, we show that $\left(f_{\rm{e}}\right)^{{N_m}}$ is an increasing and convex function of $g_m^{\rm th}$ when $g_m^{\rm th} < N_{\rm t} - 1$.

For the Q-function  ${f_{\rm Q}}\left( x \right) = \frac{1}{{\sqrt {2\pi } }}\int_x^\infty  {\exp \left( { - \frac{{{\tau ^2}}}{2}} \right)} d\tau$, we have ${f'_{\rm Q}}\left( x \right) \buildrel \Delta \over =  - \frac{1}{{\sqrt {2\pi } }}{e^{ - {x^2}/2}} < 0$, and ${f''_{\rm Q}}\left( x \right) = \frac{x}{{\sqrt {2\pi } }}{e^{ - {x^2}/2}} > 0$ when $x > 0$. Thus, ${f_{\rm Q}}\left( x \right)$ is a decreasing and convex function when $x > 0$. Since ${f_{\rm Q}}\left( x \right) < 0.5$ for $x > 0$ and $\varepsilon_m < 0.5$ that is
true for any application, and because the inverse function of a decreasing and convex function is also convex \cite{boyd}, $f_{\rm Q}^{ - 1}\left( {\varepsilon_m} \right)$ is convex in $\varepsilon_m$. Denote $z = f_{\rm Q}^{ - 1}\left( {\varepsilon_m} \right)$. Then, $g_m^{\rm th}$ in \eqref{eq:thresh} can be rewritten as follows,
\begin{align}
g_m^{{\rm{th}}} = {C_4}\left[ {\exp \left( {{C_5} + {C_6}z} \right) - 1} \right],\label{eq:gzC}
\end{align}
where $C_4 = \frac{{{N_0}{B_m}}}{{{\alpha _m}{P^{\max}}}}>0$, $C_5 = \frac{{u\ln 2}}{{T_{\rm f} {B_m}}}>0$ and $C_6 = \sqrt {\frac{1}{{T_{\rm f} {B_m}}}}>0$. It is easy to see that $g_m^{{\rm{th}}}$ is an increasing and convex function of $z$. According to the composition rules, $g_m^{\rm th}$ is convex in $\varepsilon_m$ \cite{boyd}.

From \eqref{eq:dfdg}, we can derive that
\begin{align}
\frac{{{\partial ^2}{{\left( {{f_{{\rm{e}}}}} \right)}^{N_m}}}}{{\partial {{\left( {g_m^{{\rm{th}}}} \right)}^2}}} > \frac{{{{\left( {{f_{{\rm{e}}}}} \right)}^{N_m - 1}}N_m{{\left( {g_m^{{\rm{th}}}} \right)}^{{N_{\rm{t}}} - 2}}{e^{ - g_m^{{\rm{th}}}}}}}{{\left( {{N_{\rm{t}}} - 1} \right)!}}\left( { {N_{\rm t}}} - 1 - g_m^{{\rm{th}}} \right).\nonumber
\end{align}
When $N_{\rm t} - 1 \geq g_m^{{\rm{th}}}$, ${f_{\rm{e}}}$ is increasing and convex in $g_m^{{\rm{th}}}$. According to the composition rules, ${f_{\rm{e}}}$ is convex in $\varepsilon_m$, when $N_{\rm t} - 1 \geq g_m^{{\rm{th}}}$.
\end{proof}

\bibliographystyle{IEEEtran}
\bibliography{ref}

\begin{thebibliography}{10}
\providecommand{\url}[1]{#1}
\csname url@samestyle\endcsname
\providecommand{\newblock}{\relax}
\providecommand{\bibinfo}[2]{#2}
\providecommand{\BIBentrySTDinterwordspacing}{\spaceskip=0pt\relax}
\providecommand{\BIBentryALTinterwordstretchfactor}{4}
\providecommand{\BIBentryALTinterwordspacing}{\spaceskip=\fontdimen2\font plus
\BIBentryALTinterwordstretchfactor\fontdimen3\font minus
  \fontdimen4\font\relax}
\providecommand{\BIBforeignlanguage}[2]{{%
\expandafter\ifx\csname l@#1\endcsname\relax
\typeout{** WARNING: IEEEtran.bst: No hyphenation pattern has been}%
\typeout{** loaded for the language `#1'. Using the pattern for}%
\typeout{** the default language instead.}%
\else
\language=\csname l@#1\endcsname
\fi
#2}}
\providecommand{\BIBdecl}{\relax}
\BIBdecl

\bibitem{A2014Scenarios}
{A. Osseiran, F. Boccardi and V. Braun, \emph{et al.}}, ``Scenarios for 5{G}
  mobile and wireless communications: The vision of the {METIS} project,''
  \emph{IEEE Commun. Mag.}, vol.~52, no.~5, pp. 26--35, May 2014.

\bibitem{Gerhard2014The}
G.~P. Fettweis, ``The tactile internet: Applications \& challenges,''
  \emph{IEEE Vehic. Tech. Mag.}, vol.~9, no.~1, pp. 64--70, Mar. 2014.

\bibitem{Meryem2016Tactile}
M.~Simsek, A.~Aijaz, M.~Dohler, J.~Sachs, and G.~Fettweis, ``5{G}-enabled
  tactile internet,'' \emph{IEEE J. Select. Areas Commun.}, vol.~34, no.~3, pp.
  460--473, Mar. 2016.

\bibitem{Erfan2016IoT}
E.~Soltanmohammadi, K.~Ghavami, and M.~Naraghi-Pour, ``A survey of traffic
  issues in machine-to-machine communications over {LTE},'' \emph{IEEE Internet
  of Things Journal, early access}, 2016.

\bibitem{Popovski2014METIS}
{P. Popovski, \emph{et al.}}, ``Deliverable d6.3 intermediate system evaluation
  results.''\hskip 1em plus 0.5em minus 0.4em\relax ICT-317669-METIS/D6.3,
  2014.

\bibitem{3GPP2012MTC}
G.~R1-120056, ``Analysis on traffic model and characteristics for {MTC} and
  text proposal.''\hskip 1em plus 0.5em minus 0.4em\relax Technical Report,
  TSG-RAN Meeting WG1\#68, Dresden, Germany, 2012.

\bibitem{Adnan2015Towards}
A.~Aijaz, ``Towards 5{G}-enabled tactile internet: Radio resource allocation
  for haptic communications,'' in \emph{Proc. IEEE WCNC}, 2015.

\bibitem{David2014Achieving}
D.~Ohmann, M.~Simsek, and G.~P. Fettweis, ``Achieving high availability in
  wireless networks by an optimal number of {R}ayleigh-fading links,'' in
  \emph{IEEE Globecom Workshop}, 2014.

\bibitem{Martin2015Channel}
M.~Serror, C.~Dombrowski, K.~Wehrle, and J.~Gross, ``Channel coding versus
  cooperative {ARQ}: Reducing outage probability in ultra-low latency wireless
  communications,'' in \emph{IEEE Globecom Workshop}, 2015.

\bibitem{Beatriz2015Reliable}
G.~Pocovi, B.~Soret, M.~Lauridsen, K.~I. Pedersen, and P.~Mogensen, ``Signal
  quality outage analysis for ultra-reliable communications in cellular
  networks,'' in \emph{IEEE Globecom Workshops}, 2015.

\bibitem{Yury2010Channel}
Y.~Polyanskiy, H.~V. Poor, and S.~Verd\'{u}, ``Channel coding rate in the
  finite blocklength regime,'' \emph{IEEE Trans. Inf. Theory}, vol.~56, no.~5,
  pp. 2307--2359, May 2010.

\bibitem{Gross2015Delay}
S.~Schiessl, J.~Gross, and H.~Al-Zubaidy, ``Delay analysis for wireless fading
  channels with finite blocklength channel coding,'' in \emph{Proc. ACM MSWiM},
  2015.

\bibitem{She2016GC}
C.~She, C.~Yang, and T.~Q.~S. Quek, ``Cross-layer transmission design for
  tactile internet,'' in \emph{Proc. IEEE Globecom}, 2016.

\bibitem{Johnston2015The}
R.~J. Matthew, \emph{The Role of Control Information in Wireless Link
  Scheduling}.\hskip 1em plus 0.5em minus 0.4em\relax PhD Thesis, Massachusetts
  Institute of Technology, 2015.

\bibitem{Yuan2016IoT}
Y.-C. Pang, G.-Y. Lin, and H.-Y. Wei, ``Context-aware dynamic resource
  allocation for cellular {M}2{M} communications,'' \emph{IEEE Internet of
  Things Journal}, vol.~3, no.~3, pp. 318--326, Jun. 2016.

\bibitem{Petteri2015A}
{P. Kela and J. Turkka, \emph{et al.}}, ``A novel radio frame structure for
  5{G} dense outdoor radio access networks,'' in \emph{Proc. IEEE VTC Spring},
  2015.

\bibitem{Shehzad2015Control}
S.~A. Ashraf, F.~Lindqvist, R.~Baldemair, and B.~Lindoff, ``Control channel
  design trade-offs for ultra-reliable and low-latency communication system,''
  in \emph{IEEE Globecom Workshop}, 2015.

\bibitem{Mehdi2013Performance}
M.~Khabazian, S.~Aissa, and M.~Mehmet-Ali, ``Performance modeling of safety
  messages broadcast in vehicular ad hoc networks,'' \emph{IEEE Trans. Intell.
  Transp. Syst.}, vol.~14, no.~1, pp. 380--387, Mar. 2013.

\bibitem{Yury2014Quasi}
W.~Yang, G.~Durisi, T.~Koch, and Y.~Polyanskiy, ``Quasi-static multiple-antenna
  fading channels at finite blocklength,'' \emph{IEEE Trans. Inf. Theory},
  vol.~60, no.~7, pp. 4232--4264, Jul. 2014.

\bibitem{boyd}
S.~Boyd and L.~Vandanberghe, \emph{{C}onvex {O}ptimization}.\hskip 1em plus
  0.5em minus 0.4em\relax Cambridge Univ. Press, 2004.

\bibitem{3GPP2010Shadowing}
{3GPP TR 36.814}, ``Evolved universal terrestrial radio access ({EUTRA});
  further advancements for {E}-{UTRA} physical layer aspects.''\hskip 1em plus
  0.5em minus 0.4em\relax Tech. rep. Release 9. 3GPP, 2010.

\end{thebibliography}

\end{document}